\documentclass[preprint,12pt]{elsarticle}

\usepackage{amssymb}
\usepackage{graphicx}
\usepackage{epsfig}
\usepackage{amsfonts}

\begin{document}

\begin{frontmatter}

\title{Entanglement, magnetic and quadrupole moments properties of the mixed spin Ising-Heisenberg diamond chain}

\author{V. S. Abgaryan$^{1}$}
\author{N. S. Ananikian$^{2}$}
\author{L. N. Ananikyan$^{2}$}
\author{V. Hovhannisyan$^{2}$}

\address{$^1$ Bogoliubov Laboratory of Theoretical Physics, Joint Institute for Nuclear Research, 141980 Dubna, Russia\\}
\address{$^2$ Alikhanyan National Science Laboratory, Alikhanian Br. 2, 0036 Yerevan, Armenia\\}

\begin{abstract}
Thermal entanglement, magnetic and quadrupole moments properties of
the mixed spin-$\frac{1}{2}$ and spin-1 Ising-Heisenberg model on a
diamond chain are considered. Magnetization and quadrupole moment
plateaus are observed for the antiferromagnetic couplings. Thermal
negativity as a measure of quantum entanglement of the mixed spin
system is calculated. Different behavior for the negativity is
obtained for the various values of Heisenberg dipolar and quadrupole
couplings. The intermediate plateau of the negativity has been
observed at absence of the single-ion anisotropy and quadrupole
interaction term. When dipolar and quadrupole couplings are equal
there is a similar behavior of negativity and quadrupole moment.
\end{abstract}


\end{frontmatter}

\section{Introduction}
Quantum phase transitions (QPTs)~\cite{Sach},  occurring  at a zero
temperature are triggered by changes of external parameters as a
consequence of pure quantum fluctuations. These quantum transitions
are typical for strongly correlated systems. It is known that the
system wavefunction, in general, cannot be factorized into a direct
product of subsystem states due to the nature of quantum
entanglement. This phenomenon, that does not appear in the classical
theory, is under great attention due to its importance in various
aspects of quantum information science, ranging from discrete to
continuous variable quantum computation and communication
~\cite{Nielsen,Chakhmakhchyan, zeilinger, cript}. Even more, being a
correlational measure by nature, entanglement can be a crucial
characteristic of QPTs~\cite{Wang, Hord}. The studies of entangled
chains, rings, molecules, Heisenberg models, and cluster
states~\cite{entchains1,entchains2,entrings1,entrings2}, which are
N-qubit systems, are focused on the two-qubit quantum correlations.

In the present paper we have chosen negativity ~\cite{Vidal} as a
calculable measure of entanglement, to study quantum phase
transitions and thermal entanglement in a mixed spin-($1/2$-$1$)
Ising-Heisenberg diamond chain. Thermal entanglement was detected by
both experimental~\cite{expent,Rametal,Sou} and theoretical
~\cite{Levon} observations at low dimensional spin systems, formed
in compounds.

Models on diamond chains, exhibiting interesting quantum magnetic
phenomena  have recently been  intensively studied
theoretically~\cite{Takano,CanStr,
Richter1,HidTan1,Sch,Richter3,Ananik,Richter4,Richter5,Ohanyan,
nerses}. The interest to this type of system has  been increased
since experimental work of Kikuchi and
co-workers~\cite{Kikuchi,Jesch}, where natural mineral azurite
($Cu_{3}(CO_{3})_{2}(OH)_{2}$) has been recognized as an appropriate
candidate for diamond chain compound. There have been proposed
different types of theoretical Heisenberg models to explain the
experimental measurements of magnetization plateau and the double
peak behavior in the natural mineral azurite (the density-matrix and
transfer-matrix renormalization-group techniques, density functional
theory, high-temperature expansion, variation mean-field-like
treatment, based on the Gibbs-Bogoliubov
inequality)~\cite{new,GuSu,Kang, new11}. Magnetization plateaus and
the multiple peak structure of the specific heat have also been
observed on an Ising-Hubbard diamond chain~\cite{nalb}.

The paper is organized as follows. In Section~\ref{model1} we
introduce the exactly solvable mixed spin-1/2 and spin-1
Ising-Heisenberg diamond chain with its magnetic and quadrupole
moments properties. Results for thermal negativity with and without
quadrupolar interactions and single-ion anisotropy are presented in
Section ~\ref{results}. Finally, conclusions and future prospects
are briefly mentioned in Section~\ref{conc}.

\section{Model and thermodynamics }\label{model1}

We consider the spin-$\frac{1}{2}$ and spin-$1$ Ising-Heisenberg
model on a diamond chain with quadrupole couplings  and single-ion
anisotropy (longitudinal crystal field) in the presence of the
external magnetic field. The diamond chain (Fig.~\ref{chain}) is a
quasi one-dimensional system, consisting of the nodal spins,
alternating with vertical dimers. The Hamiltonian of the model may
be written in the following form:
\begin{figure}[h!bt]
 \centerline{\psfig{figure=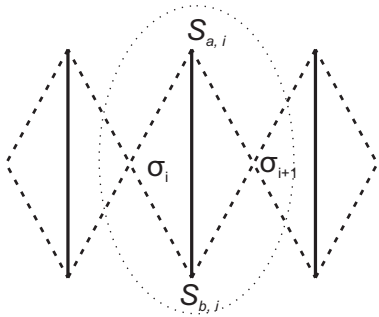,width=5 cm}}
\vspace*{8pt} \caption{A schematic representation of the mixed spin
diamond chain. Dashed (solid) lines correspond to Ising
(Heisneberg)-type interactions.}\label{chain}
\end{figure}

\begin{eqnarray}
\nonumber
 &H& = \sum _{i=1}^N H_{i},\\
 \nonumber &H_{i}&=J_{0} (\vec{S}_{a, i}\vec{S}_{b, i})+J_{1}(\sigma_{i}^{z}+\sigma_{i+1}^{z})(S_{a, i}^{z}+S_{b, i}^{z})+K(\vec{S}_{a, i}\vec{S}_{b, i})^2\\
 && +D((S_{a, i}^{z})^2+(S_{b, i}^{z})^2) -h(\frac{\sigma_{i}^{z}}{2}+S_{a, i}^{z}+\frac{\sigma_{i+1}^{z}}{2}+S_{b, i}^{z}).
\end{eqnarray}
Here $\sigma_{i}^{z}$ is Pauli $z$ matrix, $S_{a,i}^{\alpha}$ and
$S_{b,i}^{\alpha}$ ($\alpha=x,y,z$) are components of spin-1
operators. $J_{0}$ and $K$ are coefficients of dipolar and
quadrupolar interactions within spin-1 dimer, respectively. $J_{1}$
stands for Ising-type interactions between nodal spin-1/2 Ising and
spin-1 Heisenberg sites. Coefficient $D$ stands for the single-ion
anisotropy (the longitudinal crystal field) and the last term ($h$)
is the contribution of the external magnetic filed. At a special
point $J_0=K$ the system undergoes qualitative changes, namely the
operator $\sum _{i=1}^N((S_{a, i}^{z})^2+(S_{b, i}^{z})^2)$ commutes
with the total Hamiltonian and a new order parameter (quadrupole
moment) can be used to describe the properties of the system.

The important part of our further calculations is based on the
commutation relation between different block Hamiltonians
$[\mathcal{H}_i,\mathcal{H}_j]=0$ which means that the blocks of
diamond chain are separable. So in order to gain the entanglement of
the whole chain it is sufficient to calculate entanglement of single
block.

For precise solution of the model we have to consider only the
problem of the single block Hamiltonian. One may check that within
each block the Heisenberg part commutes with the Ising one. This
allows us to trace out Ising spins from the very beginning and
consider only the quantum problem of spin-1 dimer (entanglement
properties of this model are considered in ~\cite{PScripta}).
Formally, this means that we can simply replace the $\sigma_{i}^{z}$
Pauli matrixes by $\sigma_{i}=\pm 1/2$ numbers. Hence, nine
eigenvalues ($\lambda_{n}(\sigma_{i},\sigma_{i+1}), n=1,..,9$) of
the i-th block Hamiltonian can analytically be found (the
corresponding eigenvectors are also given in~\cite{PScripta}):
\begin{eqnarray}
\nonumber &&\lambda_{1,2}=J_0+2 D+K -\left(\frac{h}{2} \pm 2
J_1\right) \left(\sigma _{i}+\sigma _{i+1}\right) \pm 2 h,\\
\nonumber &&\lambda_{3,4}=J_0+ D+K -\left(\frac{h}{2} \pm
J_1\right) \left(\sigma _{i}+\sigma _{i+1}\right) \pm h,\\
\nonumber &&\lambda_{5,6}=-J_0+ D+K -\left(\frac{h}{2} \pm
J_1\right) \left(\sigma _{i}+\sigma _{i+1}\right) \pm h,\\
\nonumber &&\lambda_{7}= -J_0+2 D+K-\frac{1}{2} h \left(\sigma _{i}+\sigma _{i+1}\right),\\
\nonumber &&\lambda_{8,9}=\frac{1}{2}(-J_0+2 D+5 K-h \left(\sigma
_{i}+\sigma_{i+1}\right)\pm v),\\
\end{eqnarray}
where by $v$ we denoted:

\begin{equation}
v = \sqrt{\left(-2 D+J_0-K\right){}^2+8\left(J_0-K\right){}^2}.
\end{equation}

We are going to solve the model by direct transfer matrix method.
For this purpose let us write the partition function of the model in
this form:

\begin{equation}
Z=\sum_{\sigma_{i}}\prod^{N}_{i=1}\mbox{Tr}_{i}\rm e^{-\beta
\mathcal{H}_{i}},
\end{equation}
where $\beta=(k_{B}T)^{-1}$, $k_{B}$ is Boltzmann's constant, $T$ is
the absolute temperature. After performing a trace over the spin-1
Heisenberg dimers, one can rewrite the partition function into the
following form
\begin{eqnarray}
Z=\sum_{\sigma_{i}}\prod^{N}_{i=1}T_{\sigma_{i},\sigma_{i+1}} =
\mbox{Tr} \,T^{N},
\end{eqnarray}
here, $T_{\sigma_{i},\sigma_{i+1}}$ is the standard $2 \times 2$
transfer matrix:

\begin{eqnarray}
T_{\sigma_{i},\sigma_{i+1}} =  \left(
\begin{array}{cccc}
T_{+,+}  & T_{+,-}\\
T_{-,+} & T_{-,-}
\end{array} \right),
\end{eqnarray}

where, $\pm$ denote two spin states of the Ising spins $\sigma_i=\pm
\frac{1}{2}$. The elements of the transfer matrix are defined
through eigenvalues $(2)$ as
\begin{eqnarray}
T_{\sigma_{i},\sigma_{i+1}}=\mbox{Tr}_{i}\rm e^{-\beta
\mathcal{H}_{i}}=\sum_{n=1}^{9}\rm e^{-\beta
\lambda_{n}(\sigma_{i},\sigma_{i+1})}. \label{1.6}
\end{eqnarray}

After this, the total partition function takes the form similar to
the partition function of a one dimensional chain, with two-value
classical variables on each site:

\begin{equation}
Z=\Lambda_{1}^{N}+\Lambda_{2}^{N},
\end{equation}

where $\Lambda_{1,2}$ are the eigenvalues of transfer matrix $(6)$.
Then we take into account that in the thermodynamic limit it is
sufficient to consider only the largest eigenvalue to calculate the
partition function:
\begin{equation}
f=-\frac{1}{\beta} \mathrm{ln}\frac{1}{2}
   \left(T_{+,+}+T_{-,-}+\sqrt{\left(T_{+,+}-T_{-,-}\right){}^2+4
   T_{+,-}^2}\right).
\end{equation}

The basic order parameter for this model is the per block
magnetization, defined as
\begin{equation}
m=-\left(\frac{\partial{f}}{\partial{h}}\right)_{T,D}.
\end{equation}
As it is already mentioned above there is a special parametrization
$J_{0}=K$ with another order parameter (quadrupole moment)
\begin{equation}
q=\left(\frac{\partial{f}}{\partial{D}}\right)_{J_{0}=K,T,D}.
\end{equation}

\begin{figure}[h!]
\begin{center}
\begin{tabular}{cc}
{\small (a)}&{\small (b)}\\
\includegraphics[width=6 cm]{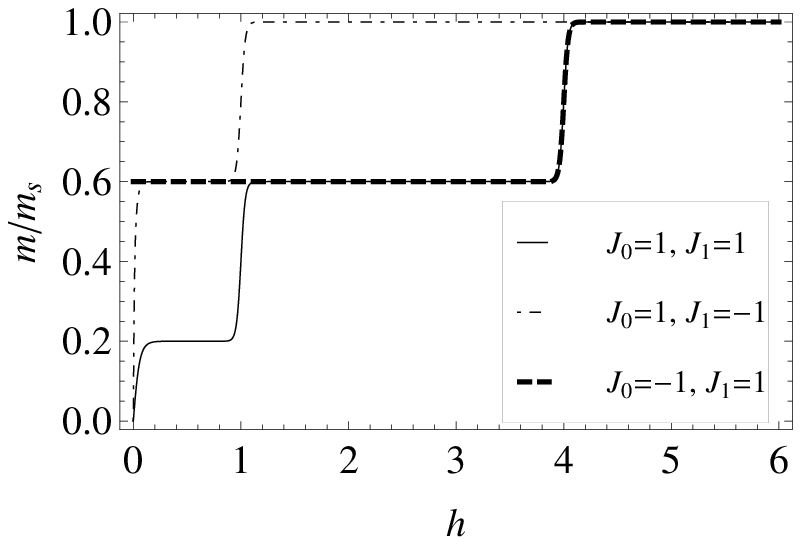}&\includegraphics[width=6 cm]{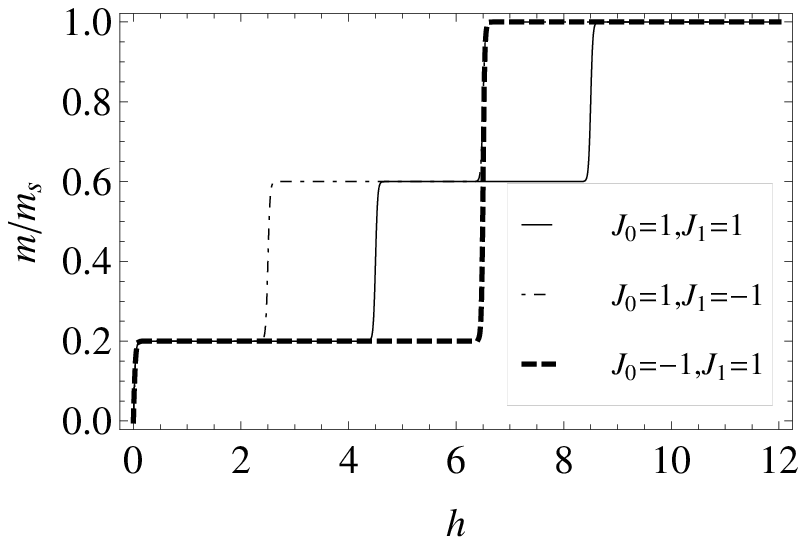}\\
 {\small(c)}&{\small(d)}\\
 \includegraphics[width=6 cm]{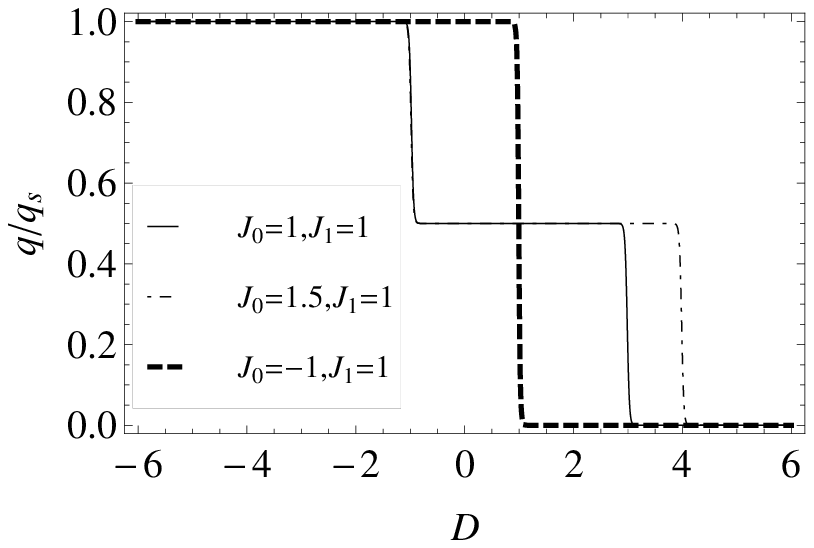}&\includegraphics[width=6 cm]{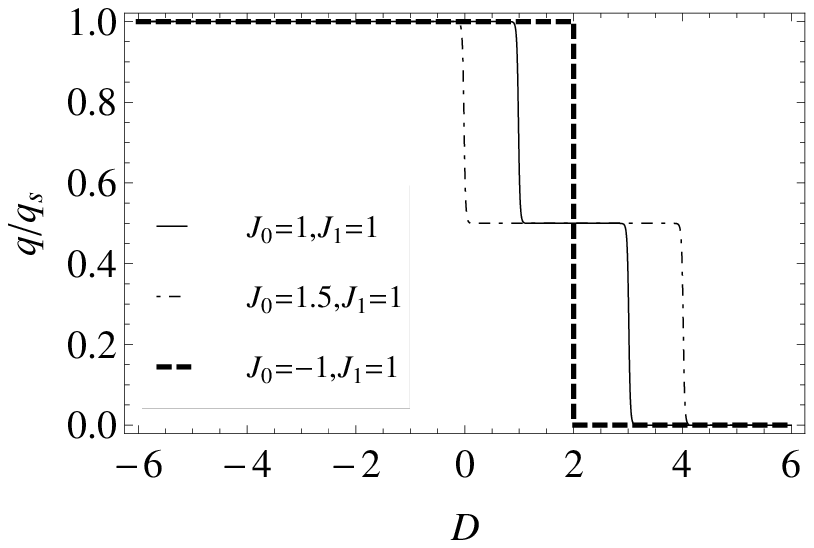}\\
\end{tabular}
\vspace*{8pt} \caption{ The magnetic-field dependence of the total
magnetization at low temperature ($T=0.02$) for different values of
exchange parameters. (a) $K=D=0$, (b) $J_0=K$ and $D=5.5$. The
single-ion anisotropy dependence of the total quadrupole moment for
different values of exchange parameters and magnetic field (c)
$h=0$, (d) $h=2$. }\label{mag}
 \end{center}
\end{figure}

General features of magnetic and quadrupole moments behavior are
presented in Fig.~\ref{mag}. First of all, let us turn our attention
to the behavior of magnetization. Figure~\ref{mag}(a) shows field
dependence of the magnetization at low temperature for the different
values of exchange couplings $J_0$ and $J_1$ ($J_{0}=1, J_{1}=1$
solid curve, $J_{0}=1, J_{1}=-1$ dot-dashed curve, $J_{0}=-1,
J_{1}=1$ dashed curve). All calculations in this case have been
carried out when quadrupolar coupling and single-ion anisotropy are
absent. The phase diagram has shown to be rather rich, demonstrating
large variety of ground states \cite{CanStr}.

The most general magnetization curve when both exchange parameters
are antiferromagnetic (solid line) shows the existence of two
intermediate plateaus at $1/5$ and $3/5$ of the saturation
magnetization. Plateau at $1/5$ corresponds to frustrated state with
dimer magnetization $\langle S^z\rangle=0$ (eigenvalues
$\lambda_{8,9}$) or state with dimer magnetization $\langle
S^z\rangle=1$, depending on the Ising spins orientation.
Magnetization plateau at $3/5$ corresponds to the dimer
magnetization $\langle S^z\rangle=1$, (more precisely to the part of
block-Hamiltonian with eigenvalues $\lambda_{5,6}$). The
magnetization reaches its saturation value at higher magnetic fields
($h>4$), thus the dimer magnetization of the ground state is
$\langle S^z\rangle=2$.

If the exchange parameters have different sign (dot-dashed and
dashed curves Fig.~\ref{mag}(a)), then, we may observe only one
intermediate plateau at $3/5$ with different values of dimer
magnetization. In the case when Heisenberg interaction is
ferromagnetic (dot-dashed curve) dimer magnetization reaches its
maximum value and remains the same for all positive values of
magnetic field. The appearance of plateau at $3/5$ is a consequence
of stepping from one set of values of classical variables (Ising
spins are parallel to the dimer spins) into another (Ising spins are
antiparallel to them). Finally, when Heisenberg interaction is
antiferromagnetic (dashed curve) transition from plateau at $3/5$ to
the saturation value takes place between states of the dimer (from
the state $\langle S^z\rangle=1$ to a state $\langle S^z\rangle=2$).
The transition between sets of values of classical variables in
particular means that entanglement of system does not change. There
are two order parameters (magnetic and quadrupole moments), when
$K=J_0$. Figure~\ref{mag}(b) shows field dependence of the
magnetization, when quadrupolar coupling and single-ion anisotropy
are presented. For all cases, one may observe the existence of the
magnetization plateau at $1/5$ in low values of the external
magnetic field.  In the case when Heisenberg interaction is
antiferromagnetic (solid and dot-dashed curves) there is an
intermediate plateau at $3/5$ with the same value of the dimer
magnetization  $\langle S^z\rangle=1$.

In Fig.~\ref{mag} (c) and (d) we plotted low-temperature quadrupole
moment curves for different values of parameters and magnetic field.
One observes only one intermediate plateau at $1/2$  when Heisenberg
interaction is antiferromagnetic (solid and dot-dashed curves). The
qualitative comparison between quadrupole moment and thermal
entanglement will be given in the next Section.

\section{ Thermal negativity}~\label{results}

In this Section, we use negativity as a computable measure of
pairwise entanglement~\cite{Vidal}, which for the density matrix
$\rho$ is defined as:

\begin{equation}
Ne(\rho)\equiv \frac{\parallel \rho^{T_1}\parallel_1 -1}{2},
 \label{neg}
\end{equation}
where $\parallel \rho^{T_1} \parallel$ is the trace norm of the
partial transposed $\rho^{T_1}$ of a bipartite density matrix
$\rho$.

At a thermal equilibrium the density matrix of the $i$-th block is
given as
\begin{equation}
{\rho}(T)=\sum_{j=1}^{36}{e^{-{E_j\over k_BT}}\over
Z_{block}}|\psi_j\rangle \langle \psi_j|,
 \label{denstymatrix}
\end{equation}
where $E_{j}$'s are eigenvalues of the block Hamiltonian $H_i$
corresponding to the states $|\psi_j\rangle$. Now let us consider
negativity properties of the mixed spin Ising-Heisenberg diamond
chain with and without quadrupolar coupling and single-ion
anisotropy.

\subsection{Chain negativity when  $K=0, D=0$}

\begin{figure}
\begin{center}
\begin{tabular}{ccc}
{\small (a)}&{\small (b)}\\
\includegraphics[width=6 cm]{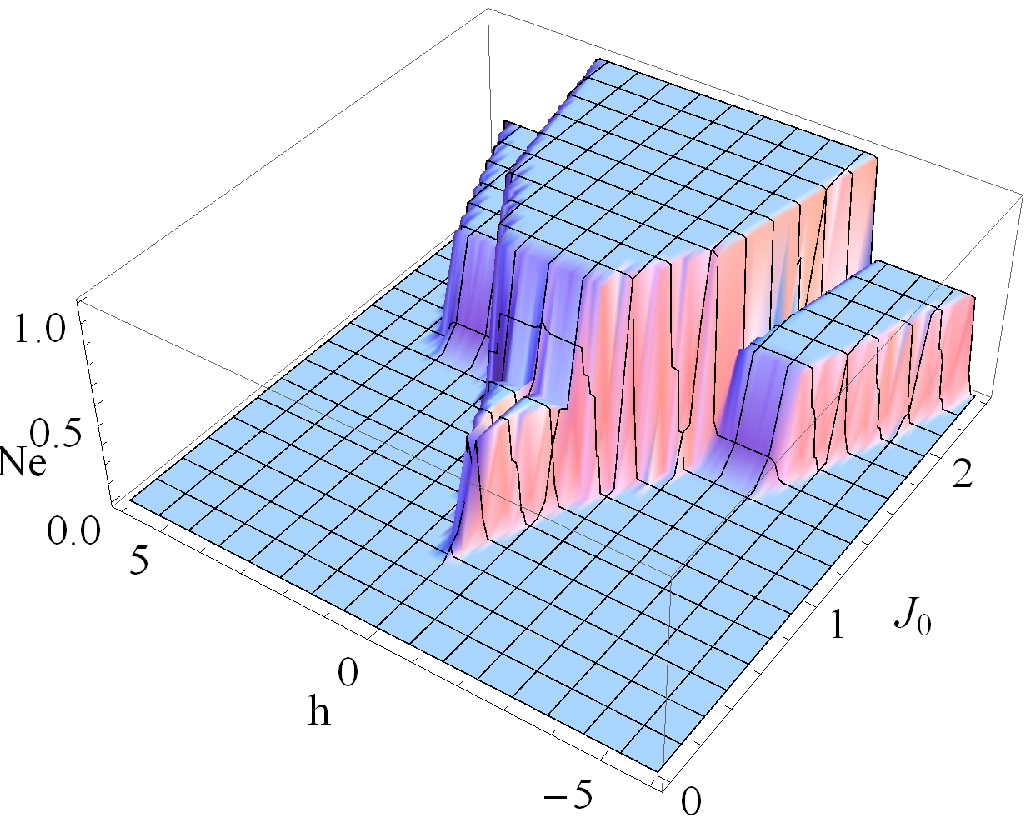}&
\includegraphics[width=6 cm]{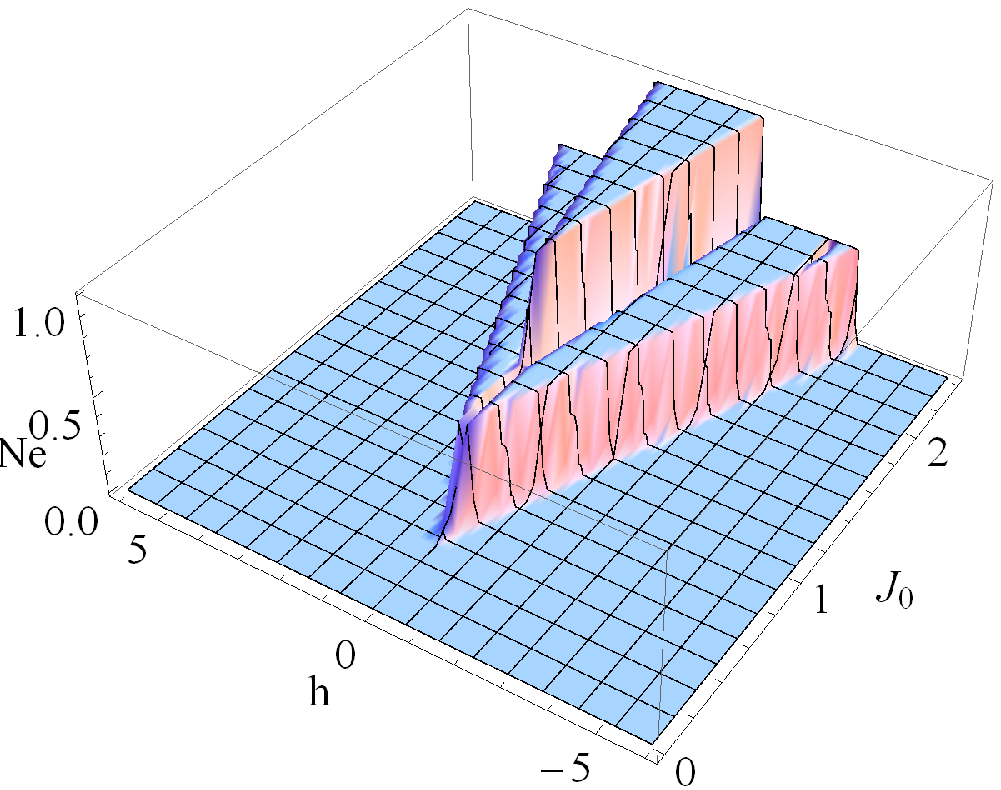}\\
\end{tabular}
\vspace*{8pt} \caption{Negativity against Heisenberg coupling
parameter $J_{0}$ and magnetic field $h$  for (a) antiferromagnetic
Ising interaction $J_{1}=1$ (b) ferromagnetic Ising interaction
$J_{1}=-1$, at low temperature ($T=0.02$) when $K=0$, $D=0$.
}\label{dens}
 \end{center}
\end{figure}

For understanding the behavior of entanglement and QPT features we
plot in Fig.~\ref{dens} a typical behavior of the negativity when
quadrupolar coupling and single-ion anisotropy are absent. Figure
\ref{dens} (a) shows magnetic field and dipolar parameter dependence
of the negativity with antiferromagnetic Ising interaction ($J_{1} =
1$). As one finds, for $0.5<J_{0}<1$ only a transition between
non-entangled $\langle S^z\rangle=2$, and partially entangled
$\langle S^z\rangle=1$ states occurs, when crossing $|h|=2J_{0}-1$
line. This transition corresponds to the magnetization jump from a
plateau at $3/5$ to $1/5$. Increasing the Heisenberg coupling
($1<J_0<1.5$), we have a transition between non-entangled and fully
entangled ($\langle S^z\rangle=0$) states that cross at
$|h|=3J_{0}-2$ line. Furthermore, for stronger Heisenberg
interaction coupling ($J_0>1.5$) the following scenario takes place.
Decreasing the absolute value of the magnetic field and crossing the
line $|h|=2J_{0}+1$, we arrive from a non-entangled state to a
partially entangled one. Finally, for even weaker magnetic fields a
transition from partially entangled state to fully entangled one
occurs at the line $|h|=2J_{0}+2$. Negativity shows a similar
behavior when Ising interaction is a ferromagnetic one (Fig.~
\ref{dens}(b)).

As for the thermal behavior of the negativity we note that it
decreases monotonically with the temperature growth, due to
decoherence effects. Particularly, for $J_{0}=\pm J_{1} $ and $h=0$
the critical temperature of entanglement vanishing is
$T_{c}=\frac{J_{0}}{\ln 2.98}\approx 0.915 J_{0}$, while for
$J_{0}\neq \pm J_{1}$ and $h=0$, $T_{c}=\frac{J_{0}-J_{1}}{\ln
2.96}\approx0.921 (J_{0}-J_{1})$.

\subsection{Negativity in special case $K=J_{0}$}

\begin{figure}[h!] \begin{center}
\begin{tabular}{cc}
{\small (a)}&{\small (b)}\\
\includegraphics[width=6 cm]{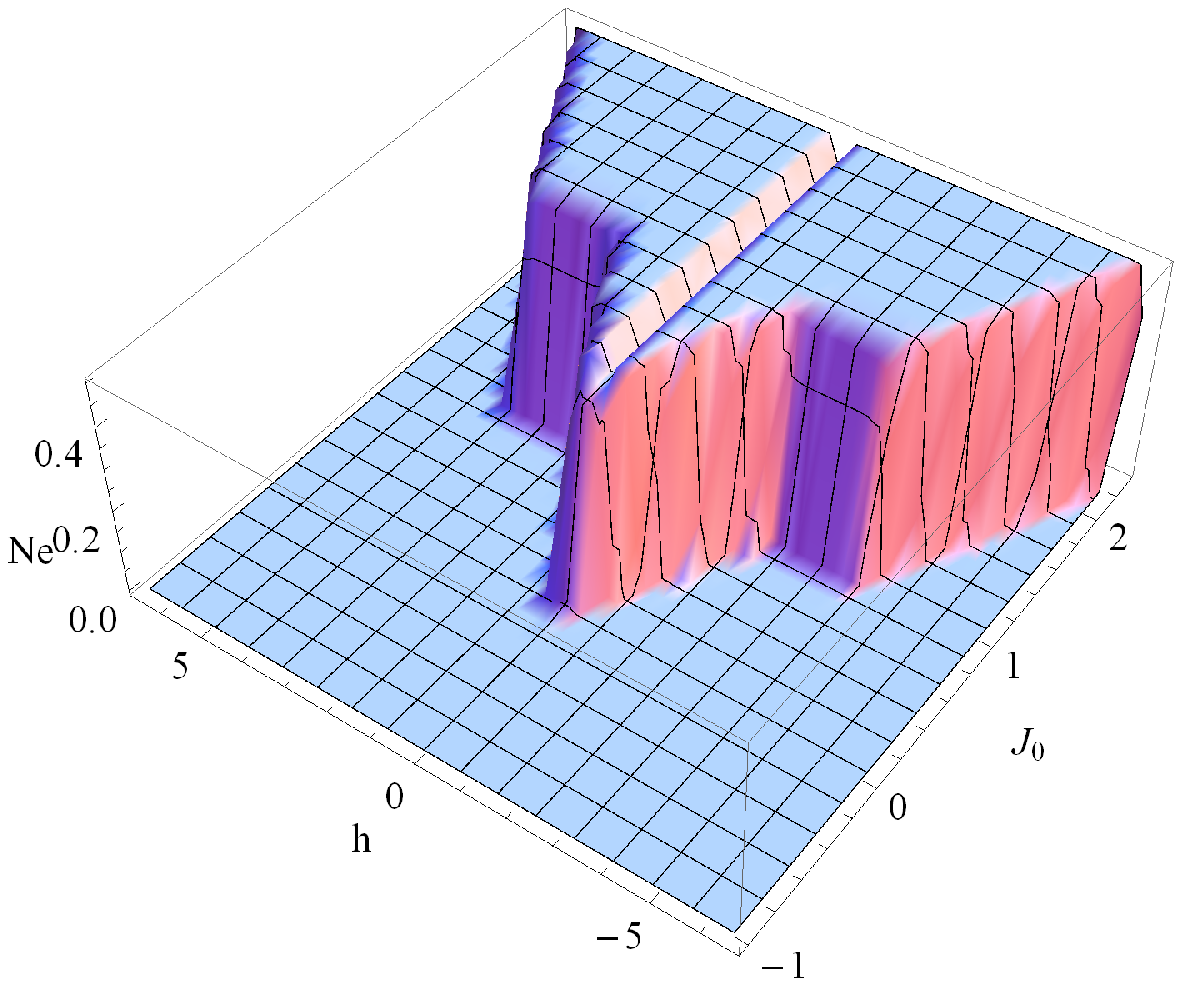}&
\includegraphics[width=6 cm]{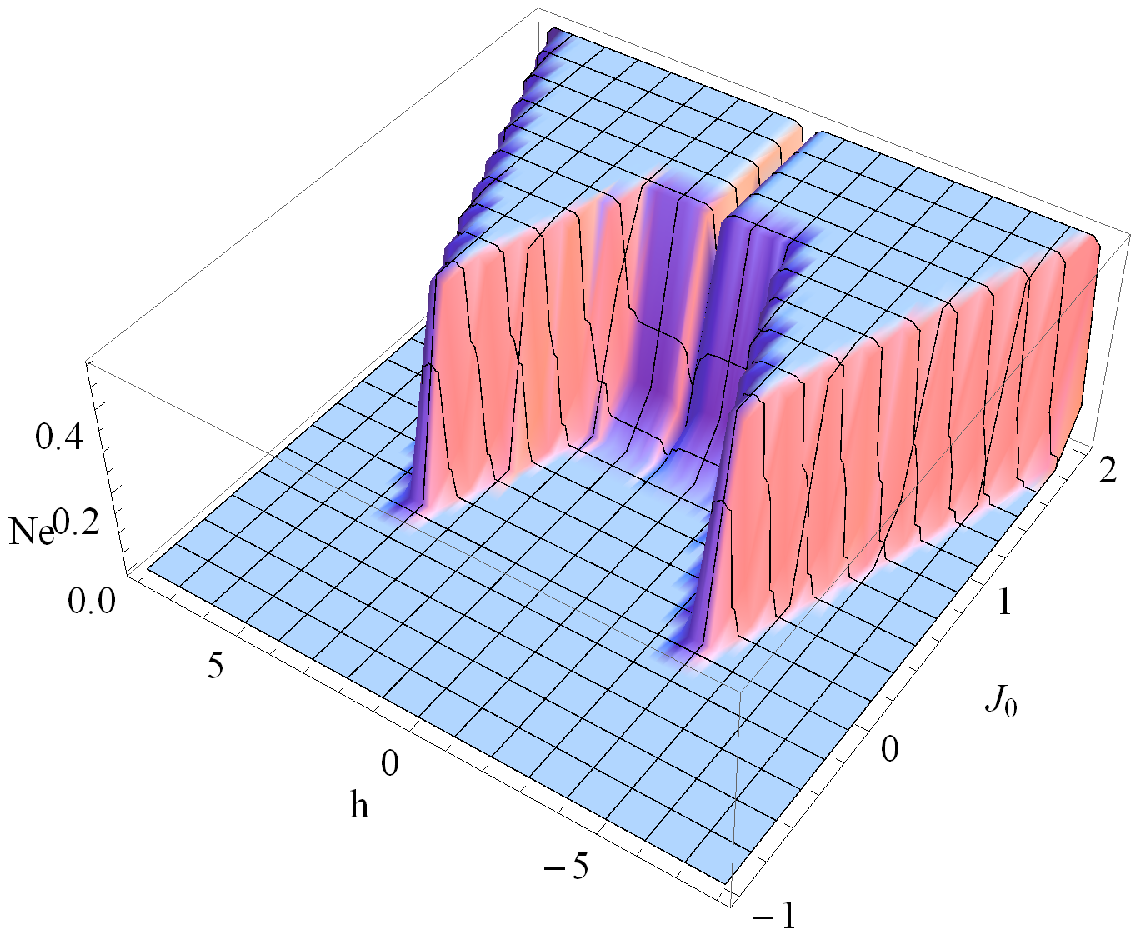}\\
\end{tabular}
\vspace*{8pt} \caption{Negativity against magnetic field and
parameter $J_{0}=K$ for the fixed value of the exchange parameter
$J_{1}=1$ and absolute temperature $T=0.1$ and for the different
values of the single-ion anisotropy (a) $D=1$, (b) $D=3$.
}\label{twographs}
 \end{center}
\end{figure}

Now let us turn our attention to the properties of the negativity
with presence of the quadrupolar coupling and single-ion anisotropy.
The typical magnetic field and dipolar parameter dependence of the
negativity is shown in Fig.~\ref{twographs} for two qualitatively
different values of the single-ion anisotropy. One can note that
$K=J_{0}$ condition brings to the disappearance of the fully
entangled states, which means that the saturation value of the
thermal negativity is 0.5.

\begin{figure}[h!] \begin{center}
\begin{tabular}{cc}
{\small (a)}&{\small (b)}\\
\includegraphics[width=6 cm]{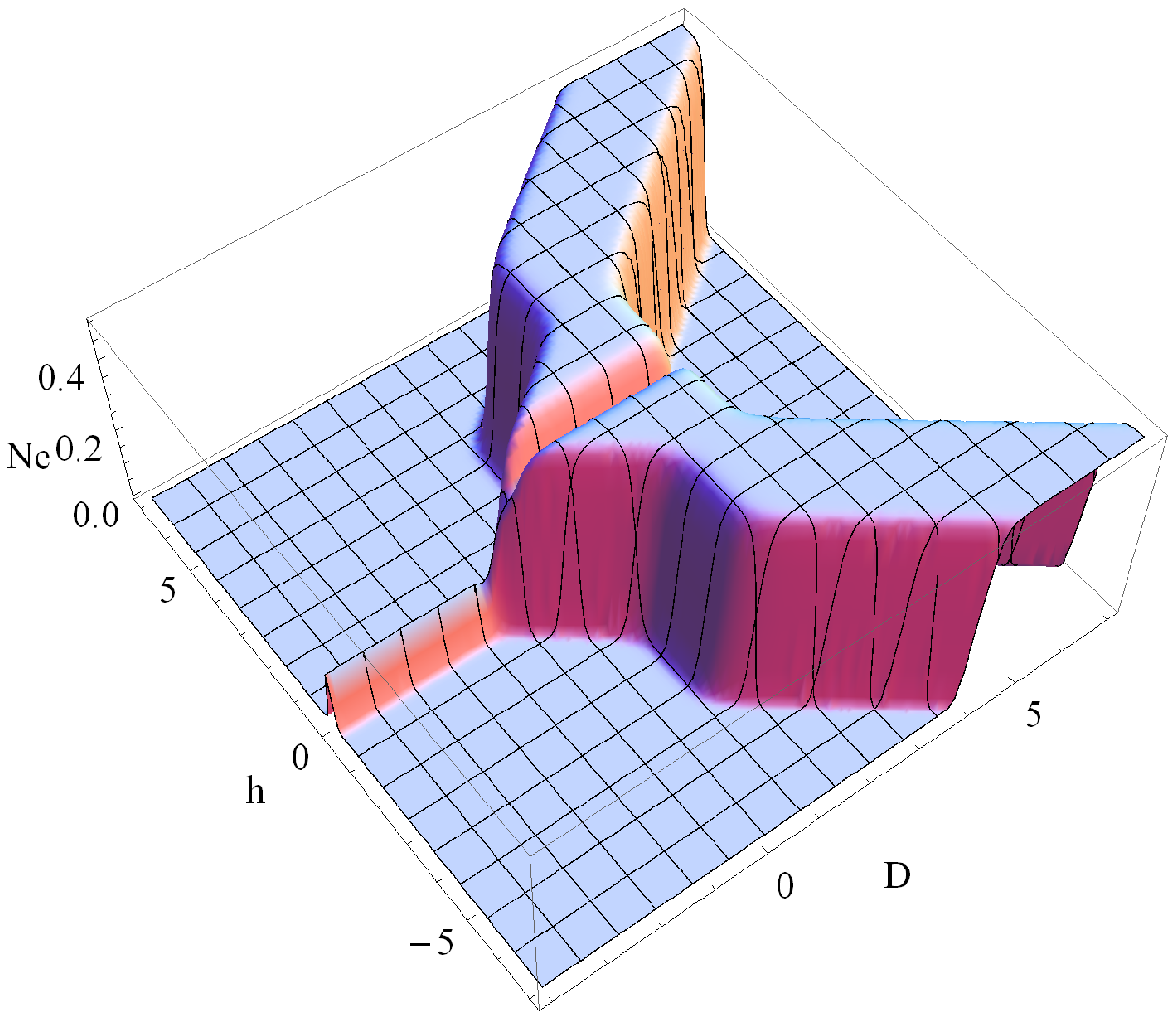}&
\includegraphics[width=6 cm]{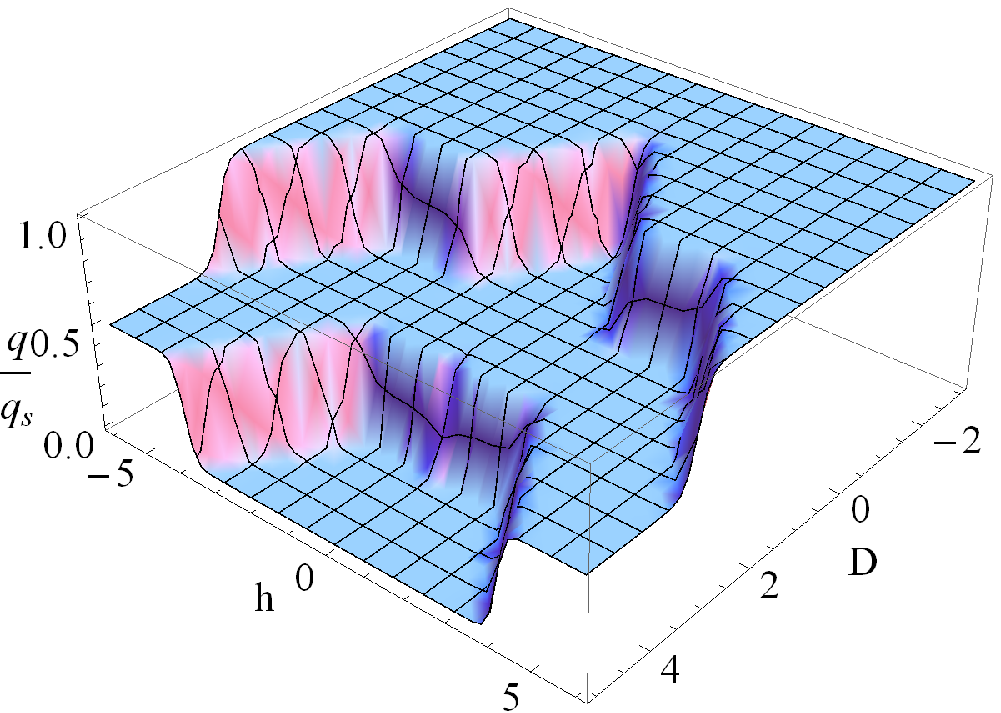}\\
\end{tabular}
\vspace*{8pt} \caption{  Negativity via magnetic field $h$ and
single-ion anisotropy $D$ when $J_{0}=J_{1}=K=1$ and $T=0.1$, (a)
quadrupole moment for same values of parameters with respect to its
saturation value (b). }\label{pp}
\end{center}
\end{figure}

Finally, let us compare properties of the thermal negativity and
quadrupole moment of the system. Figure~\ref{pp} (a) shows magnetic
field and single-ion anisotropy dependence of the negativity for the
fixed value of the exchange parameters $J_{0}=J_{1}=K=1$ and
absolute temperature $T=0.1$. As one can see, there are only two
different regimes of the negativity. Plateau at zero corresponds to
non-entangled regime, while plateau at one-half corresponds to
partially entangled one. Fig.~\ref{pp} (b) shows a three-dimensional
plot of the quadrupole moment as a function of the magnetic field
and single-ion anisotropy for the same values of the exchange
parameters and absolute temperature. One may differ three different
regions of the quadrupole moment. The first one is the region with
zero quadrupole moment, where the system is non-entangled with dimer
magnetization $\langle S^{z}\rangle=0$. Then, there is the region,
where quadrupole moment has an intermediate plateau at one-half of
the saturation value. At this region the system is partially
entangled with the maximal value of the negativity. Afterwards,
there comes a region with maximal value of the quadrupole moment
$q/q_s=1$, where the system is separable, therefore non-entangled
(with the exception of the line $h=0$).

\section{Conclusion}\label{conc}

In the present work, the mixed spin-1/2 and spin-1 Ising Heisenberg
model on a diamond chain has been exactly solved by the transfer
matrix method. In particular, we have studied the magnetic and
quadrupole moment properties of the system with and without
quadrupolar coupling and single-ion anisotropy. Particulary, we have
shown the existence of two intermediate magnetization plateaus at
one-fifth and three-fifth of the saturation magnetization. One may
also observe the existence of the plateau on the quadrupole moment
curve in the antiferromagnetic case. Using thermal negativity as a
measure of entanglement, we observe strong correlations between
magnetization, quadrupole moment and negativity. Entanglement
properties of the model strongly depend on the exchange parameters.
In particular, we obtain two or three different entangled regimes of
the system, which strongly depend on the presence of the quadrupolar
coupling and single-ion anisotropy.

\section*{Acknowledgments}
The authors would like to thank L.A. Chakhmakhchyan for useful
discussion. N A acknowledges financial support by the MC-IRSES No.
612707 (DIONICOS) under FP7-PEOPLE-2013 and research project No. SCS
13-1C137 grants. V H would like to acknowledge support from the
NFSAT, the Young Scientists Support Program (YSSP), Youth Foundation
of Armenia (AYF) and CRDF Global under grant YSSP-13-02.

\end{document}